\documentclass[a4paper,12pt]{article}
\usepackage{amssymb,amsmath}
\usepackage{bm}%
\usepackage{stmaryrd}

\def\beq{\begin{equation}}
\def\eeq{\end{equation}}
\def\bea{\begin{eqnarray}}
\def\eea{\end{eqnarray}}
\def\nn{\nonumber}

\def\i4{ {\bf 1}_4 }
\def\hf{\frac{1}{2}}
\def\del#1{\partial_{#1}}
\def\Del#1{\frac{\partial}{\partial #1}}
\def\Zgr{$ {\mathbb Z}_2$}
\def\Z2{${\mathbb Z}_2 \times {\mathbb Z}_2$}
\def\tP{ \tilde{P} }
\def\tG{ \tilde{G} }
\def\tX{ \tilde{X} }
\def\g{{\mathfrak g}}
\def\XP#1{\Pi_{#1}}
\def\XG#1{\Lambda_{#1}}
\def\F#1{\frac{\partial}{\partial #1}}
\def\gN1{ \g^{{\cal N}=1} }
\renewcommand{\theequation}{\thesection.\arabic{equation}}
\begin{document}
%
%
%
%
\thispagestyle{empty}
\vspace*{3cm}
\begin{center}
\textbf{\Large \Z2 generalizations of ${\cal N} = 2$ super Schr\"odinger algebras and their representations} \\

\vspace{10mm}
{\large
N. Aizawa${}^1$ and J. Segar${}^2$
}

\bigskip
1. Department of Physical Science, 
Graduate School of Science, 
Osaka Prefecture University, 
Nakamozu Campus, Sakai, Osaka 599-8531, Japan

\medskip
2. Department of Physics, 
Ramakrishna Mission Vivekananda College, 
Mylapore, Chennai 600 004, India

\end{center}

\vskip 3cm
\begin{abstract}
  We generalize the real and chiral $ {\cal N} =2 $ super Schr\"odinger algebras to \Z2-graded Lie superalgebras. 
This is done by $D$-module presentation and as a consequence, the $D$-module presentations of \Z2-graded superalgebras are identical to 
the ones of super Schr\"odinger algebras. We then generalize the calculus over Grassmann number to \Z2 setting. 
Using it and the standard technique of Lie theory, we obtain a vector field realization of \Z2-graded superalgebras. 
A vector field realization of the \Z2 generalization of ${\cal N} = 1 $ super Schr\"odinger algebra is also presented. 
\end{abstract}

\clearpage
\setcounter{page}{1}
%
%
%
\section{Introduction}

 Rittenberg and Wyler introduced a generalization of Lie superalgebras in 1978 \cite{rw1,rw2} (see also \cite{sch,GrJa}). 
The idea is to extend \Zgr-graded structure of Lie superalgebras to $ {\mathbb Z}_N \times {\mathbb Z}_N \times \cdots \times {\mathbb Z}_N. $ 
There are some works discussing physical applications of such an algebraic structure \cite{lr,vas,jyw,zhe,tol}, however, 
physical significance of the algebraic structure is very limited compared with Lie superalgebras. 

 Very recently, it was pointed out that a \Z2-graded Lie superalgebra (also called colour superalgebra) appears naturally as a symmetry of differential equation called 
L\'evy-Leblond equation (LLE) \cite{aktt1,aktt2}. LLE is a non-relativistic wave equation for a free, 
spin 1/2 particle whose wavefunction is a four component spinor \cite{LLE}. 
LLE is constructed so as to be  ``square root" of the free Schr\"odinger equation (similar to the relation of Dirac and Klein-Gordon equations). 
It is invariant under the Galilei group and gives the electron gyromagnetic ratio two, when coupled to a electromagnetic field.   
This shows that LLE is one of the fundamental equations in physics, though not frequently used in literatures. 
We take seriously the fact that symmetry of such a physically important equation is described by a \Z2-graded superalgebra  and decided to investigate the generalized superalgebras in more detail.

 The \Z2-graded superalgebra, which gives the symmetry of LLE, is a generalization of $ {\cal N} = 1$ super Schr\"odinger algebra (for super Schr\"odinger algebras see \cite{dh,SY1,SY2}).  
It is also shown that $ {\cal N} = 1 $ is the maximal supersymmetry of LLE  \cite{aktt1,aktt2}. 
The passage from the super Schr\"odinger algebra to its \Z2 generalization taken in Refs. \cite{aktt1,aktt2} is also applicable to other superalgebras. 
Therefore we consider in the present work  a \Z2 generalization of ${\cal N}=2$ super Schr\"odinger algebras. There are at least two inequivalent super Schr\"odinger algebra for $ {\cal N} = 2. $ 
Their $D$-module presentation is given in \cite{Dmod} and we shall use it in our construction of 
novel \Z2-graded superalgebras. As a corollary of the construction we show that the $D$-module presentations also 
give representations of the novel \Z2 superalgebras. 
These are the first main results of the present work.

  The second result is linear vector field realizations of the \Z2-graded superalgebras constructed  from $ {\cal N}=1,2$ super 
Schr\"odinger algebras. 
Symmetry generators of LLE contain second order differential operators. 
It is an interesting result that second order differential operators form a closed algebra under 
\Z2-grading, although they never close in \Zgr-grading. 
It is, however, better to have a realization of the algebras in terms of first order differential operators which would be more suitable for physical applications. 
To this end we develop a generalization of Grassmann calculus to \Z2 setting. 
Then we shall see that the techniques of standard Lie theory are used to have vector field realizations. 
These results will be applied  to built up a \Z2-graded superspace formulation. 

 The plan of this paper is as follows: The next section is a summary of already known results. 
We give a definition of \Z2-graded superalgebras and  
the $D$-module presentation of ``chiral" and ``real" super Schr\"odinger algebra with ${\cal N}=2. $ 
In \S \ref{SEC:newZ2Z2} new \Z2-graded superalgebras are constructed from ${\cal N}=2$ super Schr\"odinger algebras. We introduce a calculus over the \Z2-graded numbers in \S \ref{SEC:Number}. 
By making use of it, linear realizations of the \Z2-graded superalgebras introduced in \S \ref{SEC:newZ2Z2} are given in \S \ref{SEC:LinRep121} and \S \ref{SEC:LinRep22}. 
We also give a linear realization of \Z2-graded superalgebra obtained from ${\cal N}=1$ super 
Schr\"odinger algebra in Appendix. 

%
%
%
\setcounter{equation}{0}
\section{Preliminaries}

\subsection{\Z2-graded Lie superalgebra}

 We give a definition of \Z2-graded Lie superalgebra \cite{rw1,rw2}. 
Let $ \g $ be a linear space and $ \bm{a} = (a_1, a_2)$ an element of  \Z2. 
Suppose that $ \g $ is a direct sum of graded components:
\beq
   \g = \bigoplus_{\bm{a}} \g_{\bm{a}} = \g_{(0,0)} \oplus \g_{(0,1)} \oplus \g_{(1,0)} \oplus \g_{(1,1)}.
\eeq
Homogeneous elements of $ \g_{\bm{a}} $ are denoted by $ X_{\bm{a}}, Y_{\bm{a}}, \dots $ 
If $\g$ admits a bilinear operation (the general Lie bracket), denoted by $ \llbracket \cdot, \cdot \rrbracket, $ 
satisfying the identities:
\bea
  && \llbracket X_{\bm{a}}, Y_{\bm{b}} \rrbracket \in \g_{\bm{a}+\bm{b}}
  \\[3pt]
  && \llbracket X_{\bm{a}}, Y_{\bm{b}} \rrbracket = -(-1)^{\bm{a}\cdot \bm{b}} \llbracket Y_{\bm{b}}, X_{\bm{a}} \rrbracket,
  \\[3pt]
  && (-1)^{\bm{a}\cdot\bm{c}} \llbracket X_{\bm{a}}, \llbracket Y_{\bm{b}}, Z_{\bm{c}} \rrbracket \rrbracket
    + (-1)^{\bm{b}\cdot\bm{a}} \llbracket Y_{\bm{b}}, \llbracket Z_{\bm{c}}, X_{\bm{a}} \rrbracket \rrbracket
    + (-1)^{\bm{c}\cdot\bm{b}} \llbracket Z_{\bm{c}}, \llbracket X_{\bm{a}}, Y_{\bm{b}} \rrbracket \rrbracket =0 
    \label{gradedJ}
\eea
where
\beq
  \bm{a} + \bm{b} = (a_1+b_1, a_2+b_2) \in {\mathbb Z}_2 \times {\mathbb Z}_2, \qquad \bm{a}\cdot \bm{b} = a_1 b_1 + a_2 b_2,
\eeq
then $\g$ is referred to as a \Z2-graded Lie superalgebra. 
By definition, the general Lie bracket $ \llbracket \cdot, \cdot \rrbracket $ for homogeneous elements is either commutator or 
anticommutator. 

 This is a natural generalization of Lie superalgebra which is defined on \Zgr-grading structure:
\beq
  \g = \g_{(0)} \oplus \g_{(1)}
\eeq
with 
\beq
  \bm{a} + \bm{b} = (a+b), \qquad \bm{a} \cdot \bm{b} = ab.
\eeq
It should be noted that $ \g_{(0,0)} \oplus \g_{(0,1)} $ and $ \g_{(0,0)} \oplus \g_{(1,0)} $ are sub-superalgebras of the \Z2-graded 
superalgebra $\g.$

%
\subsection{$D$-module presentation of $ {\cal N} = 2 $ super Schr\"odinger algebras}

 We present $D$-module presentation of $ {\cal N} = 2 $ super Schr\"odinger algebras. 
For ${\cal N} = 2,$ there are two fundamental $D$-module representations of the 0+1 dimensional supersymmetry \cite{PaTo}.  
They are referred to the real and chiral representations which are also denoted as ``(1,2,1)" and ``(2,2)", respectively.  
For this reason, we have two inequivalent super Schr\"odinger algebras and we denote them $ {\cal G}_{(1,2,1)} $ and $ {\cal G}_{(2,2)}. $
Their $D$-module presentations are given below and they  correspond to $(\ell,d) = (\hf,1) $ of the ones in \cite{Dmod}. 

Let $ e_{ij} $ denote the $4 \times 4 $ supermatrix with entry $1$ at the crossing of the $i$th row and $j$th column, and $0$ otherwise. 
The $D$-module presentation of the $ {\cal G}_{(1,2,1)} $ superalgebra is given by
\bea
H &=& \i4 \del{t}, \nn \\
D &=& -\i4 \Big(t\del{t} + \lambda + \hf x \del{x} \Big) -\hf(2e_{22} + e_{33} + e_{44}),
\nn \\
K &=& -\i4 t(t\del{t} + 2\lambda + x \del{x}) -t (2e_{22} + e_{33} + e_{44}),
\nn \\
  R &=& e_{34} - e_{43},
  \nn \\
  Q_1 &=& e_{13} + e_{42} + (e_{24}+e_{31}) \del{t},
  \nn \\
  Q_2 &=& e_{14} - e_{32} - (e_{23} - e_{41}) \del{t},
  \nn \\
  S_1 &=& t(e_{13} + e_{42}) + (e_{24}+e_{31}) (t \del{t} + 2\lambda + x \del{x}) + e_{24},
  \nn \\
  S_2 &=& t(e_{14} - e_{32}) - (e_{23} - e_{41}) (t \del{t} + 2\lambda + x \del{x}) - e_{23},
  \nn \\
  P &=& \i4 \del{x},
  \nn \\
  G &=& \i4 t \del{x},
  \nn \\
  X_1 &=& (e_{24} + e_{31}) \del{x},
  \nn \\
  X_2 &=& (e_{41} - e_{23}) \del{x},  \label{G121-D-mod}
\eea
where $ \lambda $ is the constant regarded as scaling dimension.   
The \Zgr-grading structure is
\beq
  \begin{array}{ccl}
  	(0) & : & H, \ D, \ K, \ R, \ P, \ G \\[3pt]
  	(1) & : & Q_a, \ S_a, \ X_a 
  \end{array}
\eeq

  While the $D$-module presentation of the $ {\cal G}_{(2,2)} $ superalgebra is given by
\bea
  H &=& \i4 \del{t}, \nn \\
  D &=& -\i4 \Big(t\del{t} + \lambda + \hf x \del{x} \Big) -\hf(e_{33} + e_{44}),
  \nn \\
  K &=& -\i4 t(t\del{t} + 2\lambda + x \del{x}) -t (e_{33} + e_{44}),
  \nn \\
  R &=& -(e_{12} - e_{21} + e_{34} - e_{43})(x \del{x} + 2\lambda) + e_{34} - e_{43},
  \nn \\
  Q_1 &=& e_{13} + e_{24} + (e_{31} + e_{42}) \del{t},
  \nn \\
  Q_2 &=& e_{14} - e_{23} - (e_{32} - e_{41}) \del{t},
  \nn \\
  S_1 &=& t(e_{13} + e_{24}) + (e_{31} + e_{42}) (t\del{t} + 2\lambda + x \del{x}),
  \nn \\
  S_2 &=& t(e_{14} - e_{23}) - (e_{32} - e_{41}) (t\del{t} + 2\lambda + x \del{x}),
  \nn \\
  P &=& \i4 \del{x},
  \nn \\
  G &=& \i4 t \del{x},
  \nn \\   
  X_1 &=& (e_{31} + e_{42}) \del{x},
  \nn \\
  X_2 &=& -(e_{32}-e_{41}) \del{x},
  \nn \\
  J &=& (e_{12} - e_{21} + e_{34} - e_{43}) \del{x},
  \nn \\
  F &=& (e_{12} - e_{21} + e_{34} - e_{43}) t\del{x}. \label{G22-D-mod}
\eea
The \Zgr-grading structure of the superalgebra is
\beq
  \begin{array}{ccl}
  	(0) & : & H, \ D, \ K, \ R, \ P, \ G, \ J, \ F \\[3pt]
  	(1) & : & Q_a, \ S_a, \ X_a 
  \end{array}
\eeq

 For both $  {\cal G}_{(1,2,1)} $ and $  {\cal G}_{(2,2)}, $  
$ \{ H, D, K, R, Q_a, S_a \} $ spans $ sl(2|1) $ with $ \{ H, D, K \} \simeq sl(2), $ 
and $ Q_a, S_a $ supercharges and conformal supercharges, respectively. 
The rest of the elements form a $sl(2|1)$-module so that the super Schr\"odinger algebras are 
semidirect sum of $ sl(2|1) $ and its modules. 
As discussed in \cite{Dmod} (see also \cite{Naru1}), $  {\cal G}_{(1,2,1)} $ and $  {\cal G}_{(2,2)} $ accommodate a central extension. 
However, the $D$-module presentations given above are for centerless algebras.

For reader's convenience we present below nonvanishing (anti)commutators of 
$  {\cal G}_{(1,2,1)} $ and $  {\cal G}_{(2,2)}. $
First we collect the relations which are common for  
$  {\cal G}_{(1,2,1)} $ and $  {\cal G}_{(2,2)}: $
\begin{equation}
   \begin{array}{lll}
      [D, H] = H, & [D, K] = -K, & [H, K] = 2D, \\[3pt]
      [D, Q_a] = \hf Q_a, &  [K, Q_a] = S_a, & [H, S_a] = Q_a,  \\[3pt]
      [D, S_a] = -\hf S_a, & \{ Q_a, Q_b \} = 2 \delta_{ab}H, & \{ S_a, S_b \} = -2 \delta_{ab} K, \\[3pt]
      \multicolumn{2}{l}{\{ Q_a, S_b \} = - 2\delta_{ab} D + \epsilon_{ab} R,}
      \\[3pt]
     [R, Q_a] = -\epsilon_{ab} Q_b, & [R, S_a] = -\epsilon_{ab} S_b, \\[3pt]
            [H, G] = P, & [D, P] = \hf P, & [D, G] = -\hf G, \\[3pt]
       [K, P] = G, & [Q_a, G] = X_a, & [S_a, P] = -X_a,
   \end{array}
   \label{sl21plus}
\end{equation}
where $ \epsilon_{ab} $ is an antisymmetric tensor with $ \epsilon_{12} = 1 $ and summation over the repeated indices is understood.  

The superalgebra $  {\cal G}_{(1,2,1)} $ has additional relations:
\begin{equation}
  [R, X_a] = -\epsilon_{ab} X_b, \quad \{ Q_a, X_b \} = \delta_{ab} P, \quad 
       \{ S_a, X_b \} = \delta_{ab} G. 
       \label{G121add}
\end{equation}
While the additional relations for $ {\cal G}_{(2,2)} $ are given by
\begin{equation}
   \begin{array}{lll}
        \multicolumn{2}{l}{ \{Q_a, X_b\} = \delta_{ab} P - \epsilon_{ab}J,} 
        & \{ S_a, X_b \} = \delta_{ab} G - \epsilon_{ab} F,\\[3pt]
        [R, P] = J, & [R, G] = F, & [R, X_a] = -2\epsilon_{ab} X_b, \\[3pt]
        [H, F] = J, & [D, J] = \hf J, & [D, F] = -\hf F, \\[3pt]
        [K, J] = F, & [Q_a, F] = -\epsilon_{ab} X_b, & [S_a, J] = \epsilon_{ab} X_b, \\[3pt]
        [R, J]= -P, & [R, F] = -G.
   \end{array}
\end{equation}

%
%
%
\setcounter{equation}{0}
\section{\Z2 generalization of $ {\cal G}_{(1,2,1)} $ and $ {\cal G}_{(2,2)} $}
\label{SEC:newZ2Z2}

  In this section, two new \Z2-graded superalgebras are constructed from the $D$-module presentation of 
$ {\cal G}_{(1,2,1)} $ and $ {\cal G}_{(2,2)}. $ 
Key observation of the construction is that $ (Q_a, S_a), (P, G) $ are $ sl(2) $ doublet in the superalgebras. 
This suggests that one may treat $ (Q_a, S_a) $ and $ (P,G) $ on equal footing, that is, one may regard $(P,G)$ as fermionic elements. 
Doing so $ (P,G) $ generates a superalgebra and $ (Q_a, S_a) $ generate the standard superconformal algebra and interestingly enough 
their $D$-module presentation is  closed in a \Z2-graded superalgebra.

\subsection{$\g^{(1,2,1)}$ : \Z2 generalization of $ {\cal G}_{(1,2,1)} $}

Reflecting the fermionic nature of  $(P,G)$ we introduce new elements defined by 
\bea
  & & \tP = \{ P, P \}, \ U = \{P, G\}, \ \tG = \{ G, G \}, \ \Pi_a = \{P, X_a\}, \ \Lambda_a = \{G, X_a\}, 
  \nn \\
  & &  \tX = [X_1, X_2].
\eea
Note that some of them are second order differential operators in the $D$-module presentation (\ref{G121-D-mod}). 
We assign the \Z2 degree as follows:
\beq
  \begin{array}{ccl}
  	(0,0) & : & H, \ D, \ K, \ \tP, \ U, \ \tG, \ \tX, \ R \\[3pt]
  	(0,1) & : & P, \ G \\[3pt]
  	(1,0) & : & Q_a, \ S_a, \ \Pi_a, \ \Lambda_a \\[3pt]
  	(1,1) & : & X_a
  \end{array}
  \label{Z2-121}
\eeq
It is straightforward to verify, with the aid of $D$-module presentation, that the elements in (\ref{Z2-121}) satisfy the following 
nonvanishing (anti)commutation relations: 

\medskip\noindent
$(0,0)$-$(0,0)$ sector:
\beq
\arraycolsep=10pt
  \begin{array}{lll}
    [D, H] = H, & [H, K]=2D, & [D, K]=-K,  \\[3pt]
    [H, \tG]=2U, & [H, U]=\tP, & [D, \tP]=\tP,  \\[3pt]
    [D, \tG]=-\tG, & [K, \tP] = 2U,& [K, U]=\tG.
  \end{array}
\eeq

\noindent
$(0,0)$-$(0,1)$ sector:
\beq
 [H, G] = P, \qquad [D, P] = \hf P, \qquad [D, G]=-\hf G,
 \qquad [K, P] = G.
\eeq

\noindent
$(0,0)$-$(1,0)$ sector:
\beq
\arraycolsep=10pt
  \begin{array}{lll}
     [H, S_a] = Q_a, & [H, \Lambda_a] = \Pi_a, & [D, Q_a] = \hf Q_a,  \\[3pt]
     [D, S_a] = -\hf S_a, & [D, \Pi_a] = \hf \Pi_a, & [D, \Lambda_a] = -\hf \Lambda_a,  \\[3pt]
     [K, Q_a] = S_a, & [K, \Pi_a] = \Lambda_a, & [\tP, S_a] = 2 \Pi_a,  \\[3pt]
     [\tG, Q_a] = -2\Lambda_a, & [U, Q_a] = -\Pi_a, & [U, S_a] = \Lambda_a,  \\[3pt]
     [\tX, Q_a] = - \epsilon_{ab} \Pi_b, & [\tX, S_a] = -\epsilon_{ab} \Lambda_b,  \\[3pt]
     [R, {\cal X}_a ] = -\epsilon_{ab} {\cal X}_b, &
     ({\cal X} =Q, S, \Pi, \Lambda). 
  \end{array}
\eeq

\medskip\noindent
$(0,0)$-$(1,1)$ sector:
\beq
 [R, X_a] = -\epsilon_{ab} X_b.
\eeq

\noindent
$(0,1)$-$(0,1)$ sector:
\beq
  \{P, P \} = \tP, \qquad \{ P, G \} = U, \qquad \{G, G\} = \tG. 
\eeq

\noindent
$(0,1)$-$(1,0)$ sector:
\beq
  [P, S_a] = X_a, \qquad [G, Q_a] = -X_a. 
\eeq

\noindent
$(0,1)$-$(1,1)$ sector:
\beq
  \{ P, X_a \} = \Pi_a, \qquad \{ G, X_a \} = \Lambda_a.
\eeq

\noindent
$(1,0)$-$(1,0)$ sector:
\beq
\arraycolsep=10pt
  \begin{array}{ll}
    \{Q_a, Q_b\} = 2 \delta_{ab} H, & \{Q_a, S_b \} = -2 \delta_{ab} D + \epsilon_{ab} R,
    \\[3pt]
    \{ Q_a, \Pi_b \} = \delta_{ab} \tP, & \{ Q_a, \Lambda_b \} = \delta_{ab} U + \epsilon_{ab} \tX,
    \\[3pt]
     \{S_a, S_b \} = -2 \delta_{ab} K, & \{ S_a, \Pi_b \} = \delta_{ab} U - \epsilon_{ab} \tX, 
    \\[3pt]
     \{S_a, \Lambda_b \} = \delta_{ab} \tG.
  \end{array}
\eeq

\noindent
$(1,0)$-$(1,1)$ sector:
\beq
  \{ Q_a, X_b \} = \delta_{ab} P, \qquad \{ S_a, X_b \} = \delta_{ab} G. 
\eeq

\noindent
$(1,1)$-$(1,1)$ sector:
\beq
   [X_1, X_2] = \tX. 
\eeq

 It is also straightforward to verify by direct computation that the above obtained (anti)commutation relations respect the graded Jacobi identity (\ref{gradedJ}). 
We thus obtained a new \Z2-graded superalgebra, denoted as $\g^{(1,2,1)},$ and its $D$-module presentation given by (\ref{G121-D-mod}). 
 
 We used the $D$-module presentation (\ref{G121-D-mod}) to calculate the defining relations of $\g^{(1,2,1)}.$ However, the relations are representation independent. One may derive the same relations using only the (anti)commutation relations (\ref{sl21plus}) and (\ref{G121add}). 
 Thus one may realize $ \g^{(1,2,1)} $ as a subspace of the universal enveloping algebra of the ${\cal G}_{(1,2,1)}$\footnote{The authors are grateful to the referee for pointing this out.}.

%
\subsection{$\g^{(2,2)}$ : \Z2 generalization of $ {\cal G}_{(2,2)} $}

 We employ the construction same as the previous subsection. 
Introduce new elements defined by 
\bea
 & & \tP = \{ P, P \}, \ U = \{P, G\}, \ \tG = \{ G, G \}, \ V = \{P, J\}, \ W = \{P, F\}, 
 \nn \\
 & &  Z = \{ G, F \}, \ \Pi_a = \{P, X_a\}, \ \Lambda_a = \{G, X\} 
\eea
and define the \Z2-grading structure by
 \beq
 \begin{array}{ccl}
 	(0,0)  &: & H,\  D,\  K,\  \tP, \  \tG, \  U, \ V, \  W, \  Z, \  R
 	\\[3pt]
 	(0,1)  &: & P,\  G, \  J, \  F
 	\\[3pt]
 	(1,0) & : & Q_a, \  S_a,\  \Pi_a, \  \Lambda_a
 	\\[3pt]
 	(1,1) &: & X_a
 \end{array}
 \eeq
With the aid of the $D$-module presentation (\ref{G22-D-mod}) one may readily obtain the nonvanishing (anti)commutation 
relations.
 
\medskip
\noindent
$ (0,0)$-$(0,0)$ sector:
\beq
\arraycolsep=10pt
  \begin{array}{llll}
    [H, D] = -H, & [H, K] = 2D, & [H, \tG] = 2U, & [H, U] = \tP,
    \\[3pt]
    [H, W] =V, & [H, Z] = 2W, & [D, K] = -K, & [D, \tP] = \tP, 
    \\[3pt]
    [D, \tG] = -\tG, &  [D, V] = V, & [D, Z] = -Z, & [K, \tP] = 2U, 
    \\[3pt]
     [K, U] = \tG, &  [K, V] = 2W, & [K, W] =Z, & [\tP, R] = -2V,  
    \\[3pt]
    [\tG,R] = -2Z, & [U, R] = -2W, & [V, R] = 2\tP, & [W, R] = 2U,
    \\[3pt]
    [Z, R] = 2\tG.
  \end{array}
\eeq

\noindent 
$ (0,0)$-$(0,1)$ sector:
\beq
\arraycolsep=10pt
   \begin{array}{llll}
     [H,G] = P, & [H, F] = J, & [D, P] = \hf P, & [D, G] = -\hf G,
     \\[3pt]
     [D, J] = \hf J, & [D, F] = -\hf F, & [K, P] = G, & [K, J] = F,
     \\[3pt]
     [R, P] = J, & [R, G] = F, & [R, J] = -P, & [R, F] = -G.
   \end{array}
\eeq

\noindent
$ (0,0)$-$(1,0)$ sector:
\beq
\arraycolsep=10pt
   \begin{array}{llll}
     [H, S_a] = Q_a, & [H, \XG{a}] = \XP{a}, & [D, Q_a] = \hf Q_a, 
     \\[3pt]
     [D, S_a] = -\hf S_a, &  [D, \XP{a}] = \hf \XP{a}, & [D,\XG{a}] = -\hf \XG{a},
     \\[3pt]
      [K, Q_a] = S_a, & [K,\XP{a}] = \XG{a}, &  [\tP, S_a] = 2\XP{a}, 
     \\[3pt]
      [\tG, Q_a] = -2 \XG{a}, & [U, Q_a] = -\XP{a}, & [U, S_a] = \XG{a}, 
     \\[3pt]
     [V, S_a] = -2\epsilon_{ab}\XP{b}, & [W, Q_a] = \epsilon_{ab} \XP{b}, & [W, S_a] = -\epsilon_{ab} \XG{b}, 
     \\[3pt]
     [Z, Q_a] = 2 \epsilon_{ab} \XG{b}, &  [R, Q_a] = -\epsilon_{ab} Q_b, & [R, S_a] = -\epsilon_{ab} S_b, 
     \\[3pt]
     [R, \XP{a}] = -3\epsilon_{ab} \XP{b}, &   [R, \XG{a}] = -3\epsilon_{ab} \XG{b}.
   \end{array}
\eeq 
 
\noindent 
$ (0,0)$-$(1,1)$ sector:
\beq
  [R, X_a] = -2\epsilon_{ab} X_b.
\eeq

\noindent
$ (0,1)$-$(0,1)$ sector:
\beq
\arraycolsep=10pt
   \begin{array}{llll}
     \{P, P\} = \tP, & \{P, G\} = U, & \{P, J\} = V, & \{P, F\} = W,
     \\[3pt]
     \{G, G\} = \tG, & \{G, J\} = W, & \{G, F\} = Z, & \{J, J\} = -\tP,
     \\[3pt]
     \{J, F\} = -U, & \{F, F\} = -\tG.
   \end{array}
   \label{0101}
\eeq 
 
\noindent 
$ (0,1)$-$(1,0)$ sector:
\beq
  [P, S_a] = X_a, \quad [G, Q_a] = -X_a, \quad [J, S_a] = -\epsilon_{ab} X_b, \quad [F, Q_a] = \epsilon_{ab} X_b.
\eeq

\noindent
$ (0,1)$-$(1,1)$ sector:
\beq
  \{P, X_a \} = \XP{a}, \quad \{G, X_a\} = \XG{a}, \quad \{J, X_a\} = -\epsilon_{ab} \XP{b}, \quad 
  \{F, X_a\} = -\epsilon_{ab} \XG{b}.
\eeq
 
\noindent 
$ (1,0)$-$(1,0)$ sector:
\beq
\arraycolsep=10pt
   \begin{array}{ll}
     \{Q_a, Q_b\} = 2 \delta_{ab}H, & \{Q_a, S_b\} = -2\delta_{ab} D + \epsilon_{ab} R, 
     \\[3pt]
     \{Q_a, \XP{b}\} = \delta_{ab} \tP - \epsilon_{ab} V, & 
     \{Q_a, \XG{b}\} = \delta_{ab} U - \epsilon_{ab} W,
     \\[3pt]
     \{S_a, S_b\} = - 2 \delta_{ab} K, & \{ S_a, \XP{b} \} = \delta_{ab} U - \epsilon_{ab} W,
     \\[3pt]
     \{ S_a, \XG{b} \} = \delta_{ab} \tG - \epsilon_{ab} Z.
   \end{array}
\eeq

\noindent
$ (1,0)$-$(1,1)$ sector:
\beq
 \{ Q_a, X_b \} = \delta_{ab} P - \epsilon_{ab}J, \qquad 
 \{ S_a, X_b \} = \delta_{ab} G - \epsilon_{ab}F.
\eeq
There is no nonvanishing relations in $ \g_{(1,0)}$-$\g_{(1,1)}$ sector.  
One may verify by direct computation that the above obtained (anti)commutation relations respect the graded Jacobi identity. 
We thus obtained a new \Z2-graded superalgebra, denoted as $ \g^{(2,2)},$ and its $D$-module presentation given by (\ref{G22-D-mod}).

Contrary to $ \g^{(1,2,1)}$ it is impossible to realize $ \g^{(2,2)}$ as a subspace of the universal enveloping algebra of $ {\cal G}_{2,2}$ 
since the relations in (\ref{0101}) are never realized in the enveloping algebra. 

%
%
%
\setcounter{equation}{0}
\section{\Z2 generalization of Grassmann numbers and their calculus}
\label{SEC:Number}

 The $D$-module presentation of $ \g^{(1,2,1)} $ and $ \g^{(2,2)} $ discussed in \S \ref{SEC:newZ2Z2} consists of first and second order 
differential operators. Appearance of second order differential operators in a representation of Lie (super)algebra is unusual in physics. 
Concerning physical applications of $ \g^{(1,2,1)} $ and $ \g^{(2,2)} $ it would be more appropriate to have a representation of them 
consisting of only first order differential operators. 
This would be achieved by generalizing the Grassmann calculus to \Z2 setting. 
We, therefore, develop it in this section.

  A \Z2 generalization of Grassmann numbers was introduced in \cite{rw1,rw2}. 
\Z2 graded numbers $\xi_{\bm{a}}, \xi_{\bm{b}} $ with degree $ \bm{a}, \bm{b} \in$ \Z2 are defined by the relation:
\beq
   \llbracket \xi_{\bm{a}}, \xi_{\bm{b}} \rrbracket =0. \label{Z2GrassDef}
\eeq
It would be more helpful to use different notations for numbers with different degree. 
We denote the \Z2 graded numbers as follows:
\beq
 (0,0) \ x_i, \quad (0,1) \ \psi_i, \quad (1,0) \ \theta_i, \quad (1,1)\ z_i 
\eeq
where their \Z2 degree is also indicated. 
By definition (\ref{Z2GrassDef})  
$ x_i $ commute with all others and  $ \psi_i^2 = \theta_i^2 = 0. $ 
$ z_i $ anticommute with $ \psi_i, \theta_i, $ however, $ \psi_i $ and $ \theta_i $ commute each other. 
The simplest example is the system consisting of $ 1, x, \psi, \theta $ and $z.$ 
A function of these variables is expanded in powers of nilpotent variables $ \psi, \theta:$
\beq
  f(x,\psi,\theta,z) = f_{00}(x,z) + \psi f_{01}(x,z) + \theta f_{10}(x,z) + \psi \theta f_{11}(x,z). 
  \label{f-simplest}
\eeq
Note that the function $ f_{ab}$ has the degree $(a,b). $

  Now we define derivative and integral for these numbers in such a way the definitions are 
a  natural generalization of the Grassmann case. 
First, $ \displaystyle \F{A} $ and $ dA $ are required to have the degree same as $ A = x, \psi, \theta, z $ so that they obey the (anti)commutation relations same as $ A. $  
We treat $x_i$  as an ordinary real (or complex) number, thus its derivative and integral are obvious. 
$ z_i $ is also treated as an ordinary number except that it anticommutes with $ \psi_i $ and $ \theta_i. $  
Derivative and integral for $ \psi_i $ and $ \theta_i $ are same as Grassmann case:
\bea
  & & \frac{\partial}{\partial\psi_i} = \int d\psi_i, \qquad \frac{\partial}{\partial\theta_i} = \int d\theta_i,
  \nn \\
  & & \frac{\partial}{\partial\psi_i} 1 = \frac{\partial}{\partial\theta_i} 1 = 0, 
  \qquad
   \frac{\partial}{\partial\psi_i} \psi_j = \frac{\partial}{\partial\theta_i} \theta_j = \delta_{ij}.
\eea 
To clarify our definition, we here give some examples:
\bea
  & & 
  \frac{\partial}{\partial\psi_2} x_2 \psi_1 \psi_2 = - x_2 \psi_1, 
  \qquad\qquad
  \frac{\partial}{\partial\theta_1} \psi_1 \theta_1 z_3 = \psi_1 z_3,
  \\
  & &
  \frac{\partial}{\partial z_1} x_2 \psi_3 z_1^2 = -2 x_2\psi_3 z_1,
\eea
This definition guarantees that the integration by parts holds for these numbers.

%
%
%
\setcounter{equation}{0}
\section{Linear realizations of $\g^{(1,2,1)}$}
\label{SEC:LinRep121}

The aim of this section is to realize $\g^{(1,2,1)}$ by first order differential operators of \Z2-graded numbers. 
Procedure employed here is a standard one in Lie theory. We apply it to our \Z2 setting.   
Namely, for a \Z2-graded superalgebra we shall consider triangular-like decomposition, exponential mapping, functions over a Lie group, \textit{etc}.  

\subsection{Triangular-like decomposition of $\g^{(1,2,1)}$}

We would like to introduce a vector space decomposition of $\g^{(1,2,1)}$ similar to the triangular decomposition of a Lie algebra. 
To this end, we introduce another grading for $\g^{(1,2,1)}. $ 
Taking $D$ as a grading operator, $D$ degree $ \delta $ of $ Y \in \g^{(1,2,1)} $ is defined by 
$ [D, Y] = \delta Y. $  
We also introduce the second grading operator $ \bar{R} = -i R $ and its degree $ \rho $ which is defined the way same as $\delta.$  
Pair of $D$ and $ \bar R $ degree $(\delta,\rho)$ for each element of $ \g^{(1,2,1)} $ follows immediately from the defining (anti)commutation 
relations:
 \beq
    \begin{array}{llllll}
     H (1,0), & D (0,0), & K (-1,0), & \tP (1,0), & U (0,0), & \tG (-1,0), \\[3pt]
     \tX (0,0), & \bar{R} (0,0), & P (\hf,0), & G (-\hf,0), & X_{\pm} (0, \pm 1), & Q_{\pm} (\hf, \pm 1), \\[3pt]
     S_{\pm} (-\hf, \pm 1), & \Pi_{\pm} (\hf, \pm 1), & \Lambda_{\pm} (-\hf, \pm 1)
    \end{array}
 \eeq 
where $ {\cal X}_{\pm} = {\cal X}_1 \pm i {\cal X}_2 $ for $ {\cal X} = X, Q, S, \Pi, \Lambda. $  
Reading the entries of the pair from left to right, if the first nonvanishing 
entry is positive (negative) then we call the element of $ \g^{(1,2,1)} $ positive (negative). 
According to this, we introduce a vector space decomposition of $ \g^{(1,2,1)} $ similar to the triangular decomposition:
\beq
   \g^{(1,2,1)} = \g_+^{(1,2,1)} \oplus \g_0^{(1,2,1)} \oplus \g_-^{(1,2,1)},
\eeq
with
\bea
  \g_+^{(1,2,1)} &:& H, \ \tP, \ P, \  \ Q_{\pm}, \ \Pi_{\pm}, \ X_+ 
  \nn \\
  \g_0^{(1,2,1)} &:& D, \ U, \ \tX, \ \bar{R}
  \nn \\
  \g_-^{(1,2,1)} &:& K, \tG, \ G, \ S_{\pm}, \ \Lambda_{\pm}, \ X_-
\eea
One may easily see that
\beq
 [\g_0^{(1,2,1)}, \g_{\pm}^{(1,2,1)}] \subseteq \g_{\pm}^{(1,2,1)}. 
\eeq

  By exponential mapping the \Z2-graded supergroup is obtained from $ \g^{(1,2,1)} $ 
\beq
   G^{(1,2,1)}  = \exp(\g^{(1,2,1)} )
\eeq
and the corresponding triangular-like decomposition:
\beq
  G^{(1,2,1)} = G^{(1,2,1)}_+ G^{(1,2,1)}_0 G^{(1,2,1)}_-, 
\eeq
where
\beq
  \qquad G^{(1,2,1)}_{\pm} = \exp(\g^{(1,2,1)}_{\pm}), \quad G^{(1,2,1)}_0 = \exp(\g^{(1,2,1)}_0).
\eeq

%
\subsection{Left action of $ \g^{(1,2,1)} $ on $G^{(1,2,1)}$}

We consider the space of $ C^{\infty} $ class functions over $G^{(1,2,1)}$ with a special property defined by
\beq
  f(g_+ g_0 g_-) = f(g_+), \quad g_{a} \in G^{(1,2,1)}_a, \ a = 0, \pm
\eeq
Namely, $f$ is a function over $G^{(1,2,1)}_+.$
We parametrize elements of $ G^{(1,2,1)}_+ $ as follows:
\bea
  g_+ &=& \exp(x_1 H) \exp(x_2 \tP) \exp(\psi P) \exp(\theta_{+} Q_{+}) \exp(\theta_{-} Q_{-})
  \nn \\[3pt]
 && \qquad \times
   \exp(\theta'_{+} \Pi_{+}) \exp(\theta'_{-} \Pi_{-}) \exp(z X_{+}).
\eea
We define a left action of $\g^{(1,2,1)}$ on $ f(g_+) $ as usual:
\beq
   Y f(g_+) = \left. \frac{d}{d\tau} f(e^{-\tau Y}g_+) \right|_{\tau=0}, \quad Y \in \g^{(1,2,1)} \label{LAdef}
\eeq
where $ \tau $ is a parameter having the degree same as $ Y.$ 
It is not difficult to verify that the left action gives a realization of $ \g^{(1,2,1)} $ on the space of functions $f(g_+). $ 
We give its explicit form  below:
  
\medskip
\noindent
$\g^{(1,2,1)}_+$ sector:
\bea
  H &=& -\F{x_1},
  \nn \\
  \tP &=& -\F{x_2},
  \nn \\
  P &=& -\F{\psi} + \hf \psi \F{x_2}, 
  \nn \\
  Q_+ &=&  -\F{\theta_+},
  \nn \\
  Q_- &=& -\F{\theta_-} + 4 \theta_+ \F{x_1},
  \nn \\
  \Pi_{\pm} &=& - \F{\theta'_{\pm}} + 2 \theta_{\mp} \F{x_2},
  \nn \\
  X_+ &=& -\F{z} + 2\theta_- \F{\psi} - \psi \theta_- \F{x_2} + \psi \F{\theta_+'}.
\eea

\noindent
$\g^{(1,2,1)}_0$ sector:
\bea
  D &=& - \sum_{a=1,2} x_a \F{x_a} - \hf \sum_{k=\pm} \Big( \theta_k \F{\theta_k} + \theta'_k \F{\theta'_k} \Big) - \hf \psi \F{\psi},
  \nn \\
  U &=& (x_1- 2\theta_+ \theta_-) \F{x_2} + \theta_+ \F{\theta'_+} + \theta_- \F{\theta'_-},
  \nn \\
  \tX &=& i \Big( 2\theta_+ \theta_- \F{x_2} -\theta_+ \F{\theta'_+} + \theta_- \F{\theta'_-}\Big),
  \nn \\
  \bar{R} &=& -\sum_{k} \text{sign}(k) \Big( \theta_k \F{\theta_k} + \theta'_k \F{\theta'_k} \Big) -z \F{z},
\eea
withe sign$(\pm) = \pm 1. $

\noindent
$\g^{(1,2,1)}_-$ sector:
\bea
  K &=& -2x_1 D -x_1^2 \F{x_1}  - 2\theta_+ \theta_- \Big(2 \theta'_+ \F{\theta'_+} + 2 x_2 \F{x_2} + \psi \F{\psi} + z \F{z}  \Big) 
  \nn \\
  &+& 2x_2\Big( \theta_+ \F{\theta'_+} + \theta_- \F{\theta'_-} \Big) + \psi \theta_+ \F{z},
  \nn \\
  \tG &=& -x_1 \Big( U - \theta_+ \Pi_+ + \theta_- \F{\theta'_-} \Big),
  \nn \\
  G &=& -x_1 P - \theta+ X_+ - \psi \theta_- \F{\theta'_-},
  \nn \\
  S_+ &=& -x_1 Q_+ - 2 x_2 \Pi_+ - 2\theta_- \Big( \psi \F{\psi} + 2 \theta'_+ \F{\theta'_+} + z \F{z}  \Big) + \psi \F{z},
  \nn \\
  S_- &=& -x_1 Q_- - 2 x_2 \Pi_- - 2\theta_+ \Big( \psi \F{\psi} + 2\theta_- \F{\theta_-} + 2 \theta'_- \F{\theta'_-} - z \F{z}  \Big),
  \nn \\
  \Lambda_+ &=& -x_1 \Pi_+,
  \nn \\
  \Lambda_- &=& -x_1 \Pi_- - 4\theta_+ \theta_- \F{\theta'_-},
  \nn \\
  X_- &=& 2\theta_+ \F{\psi} - \psi \theta_+ \F{x_2} + \psi \F{\theta'_-}.
\eea

%
%
%
\setcounter{equation}{0}
\section{Linear realizations of $\g^{(2,2)}$}
\label{SEC:LinRep22}

  By the procedure same as the one for $ \g^{(1,2,1)} $ we construct a linear realization of $ \g^{(2,2)}. $ 
First, we redefine the elements:
\bea
   & & V_{\pm} = V \pm i \tP, \quad Z_{\pm} = Z \pm i \tG, \quad W_{\pm} = W \pm i U, 
   \nn \\
   & & J_{\pm} = J \pm i P, \quad F_{\pm} = F \pm i G.
\eea
Taking $D$ and $ \bar{R} = -i R $ as the grading operators, the $(\delta, \rho) $ degree of  $\g^{(2,2)} $ elements is given by
\beq
   \begin{array}{llllll}
      H (1,0), & D (0,0), & K (-1,0), & V_{\pm} (1, \pm 2), & W_{\pm} (0, \pm 2), & Z_{\pm} (-1, \pm 2), \\[3pt]
      \bar{R} (0,0), & J_{\pm} (\hf,\pm 1), & F_{\pm} (-\hf, \pm 1), & Q_{\pm} (\hf, \pm 1), & S_{\pm} (-\hf, \pm 1), &
      \Pi_{\pm} (\hf, \pm 3), \\[3pt]
      \Lambda_{\pm} (-\hf, \pm 3), & X_{\pm} (0, \pm 2)
   \end{array}
\eeq
This leads a triangular-like decomposition of $ \g^{(2,2)} $
\bea
 \g^{(2,2)}_+ &:& H, \ V_{\pm}, \ W_+, \ Q_{\pm}, \ J_{\pm}, \ \Pi_{\pm}, \ X_+
 \nn \\
 \g^{(2,2)}_0 &:& D, \ \bar{R}
 \nn \\
 \g^{(2,2)}_- &:& K, \ Z_{\pm},\ W_-, \ S_{\pm}, \ F_{\pm}, \ \Lambda_{\pm}, \ X_-
\eea
We parametrize an element of $ G^{(2,2)}_+ = \exp(\g^{(2,2)}_+) $  
as follows:  
\bea
  g_+ &=& \exp(x_1 H) \exp(x_2 W_+) \exp(x_+ V_+) \exp(x_- V_-) \exp(\psi_+ J_+) \exp(\psi_- J_-)
  \nn \\[3pt]
  && \qquad \times \exp(\theta_+ Q_+) \exp(\theta_- Q_-) \exp(\theta'_+ \Pi_+) \exp(\theta'_- \Pi_-)
  \exp(z X_+).
\eea
Then the left action (\ref{LAdef}) of $\g^{(2,2)} $ on the space of functions over $G^{(2,2)}_+ $ 
gives a linear realization of $ \g^{(2,2)}. $ 
We give its explicit form  below:

\medskip
\noindent
$\g^{(2,2)}_+$ sector:
\begin{eqnarray}
  H &=& -\F{x_1},
  \nn \\
  W_+ &=& -\F{x_2} + x_1 \F{x_+},
  \nn \\
  V_{\pm} &=& - \F{x_{\pm}},
  \nn \\
  J_{\pm} &=& -\F{\psi_{\pm}} \pm i \psi_{\pm} \F{x_{\pm}},
  \nn \\
  Q_+ &=& -\F{\theta_+} + 2x_2 \Big( 2\theta_- \F{x_+} - i \F{\theta'_+} \Big),
  \nn \\
  Q_- &=& -\F{\theta_-} - 4 \theta_+ \Big( x_2 \F{x_+} - \F{x_1} \Big), 
  \nn \\
  \Pi_{\pm} &=& -\F{\theta_{\pm}'} \mp 2i \theta_{\mp} \F{x_{\pm}},
  \nn \\
  X_+ &=& -\F{z} -2 \Big( \psi_+ \theta_- \F{x_+} + i \theta_- \F{\psi_+} - i \psi_+ \F{\theta'_+} \Big).  
\end{eqnarray}

\noindent
$\g^{(2,2)}_0$ sector:
\begin{eqnarray}
 D &=& -x_1 \F{x_1} -\sum_{k=\pm} \left( x_k \F{x_k} + \frac{1}{2} \Big( \psi_k \F{\psi_k} + \theta_k \F{\theta_k} + \theta'_k \F{\theta'_k}\Big) \right),
 \nn \\
 \bar{R} &=& -2x_2 \F{x_2} - \sum_{k=\pm} \text{sign}(k)  \Big( 2x_k \F{x_k} + \psi_k \F{\psi_k} + \theta_k \F{\theta_k} + 3 \theta'_k \F{\theta'_k} \Big) - 2 z \F{z}. 
\end{eqnarray}

\noindent
$\g^{(2,2)}_-$ sector:
\begin{eqnarray}
   W_- &=& -2i \theta_- \F{\theta'_-} + x_1 \F{x_-},
   \nn \\
   Z+ &=& 2(x_1 - 4\theta_+ \theta_-) \F{x_2} - x_1^2 \F{x_+},
   \nn \\
   Z_- &=& 4i x_1 \theta_- \F{\theta'_-} - x_1^2 \F{x_-},
   \nn \\
   X_- &=& -2 \Big( \psi_- \theta_+ \F{x_-} - i\theta_+ \F{\psi_-} + i \psi_- \F{\theta'_-} \Big),
   \nn \\
   F_+ &=& -x_1 J_+ + 2i \Big(  \psi_+ \F{x_2} - 2\psi_+ \theta_+ \theta_- \F{x_+} + 2i \theta_+ \theta_- \F{\psi_+} - \theta_+ \F{z} \Big), 
   \nn \\
   F_- &=& -x_1 J_- + 4 \psi_- \theta_- \F{\theta'_-},
   \nn \\
   \Lambda_+ &=& -x_1 X^P_+ + 2\theta_- \Big( 2\theta_+ \F{\theta'_+} - i \F{x_2} \Big),
   \nn \\
   \Lambda_- &=& -x_1 X^P_- -4\theta_+ \theta_- \F{\theta'_-},
   \nn \\
   S_+ &=& -x_1 Q_+ -4i x_+ X^P_+ 
   -4 \theta_- \Big( x_2 \F{x_2} + \psi_+ \F{\psi_+} + 2 \theta'_+ \F{\theta'_+} - \theta'_- \F{\theta'_-} + z \F{z}  \Big)
   \nn \\
   &+& 2i \Big( 4x_2 \theta_+ \theta_- \F{\theta_+'} + \psi_+ \F{z} \Big),
   \nn \\
   S_- &=& -x_1 Q_- +4i x_- X^P_- 
   -4\theta_+ \Big( \psi_- \F{\psi_-} + \theta_- \F{\theta_-} + 2 \theta'_- \F{\theta_-'} - z\F{z}  \Big)
   \nn \\
   &-& 2i \theta_+' \Big(  \F{x_2}  - 4 \theta_+ \theta_- \F{x_+} \Big),
   \nn \\
   K &=& -2x_1 D -x_1^2 \F{x_1} -4 \theta_+  \theta_- 
   \Big( 2x_2 \F{x_2} + \psi_+ \F{\psi_+} + \theta_+' \F{\theta_+'} - \theta_-' \F{\theta_-'} + z \F{z}  \Big) 
   \nn \\
   &-&2(x_+ + i \theta_- \theta_+') \F{x_2} - 4ix_- \theta_- \F{\theta_-'} 
   +2 i \psi_+ \theta_+ \F{z}.
\end{eqnarray}   
   
%
%
%
\section{Concluding remarks}

  We have introduced two novel \Z2-graded Lie superalgebras $ \g^{(1,2,1)} $ and $ \g^{(2,2)} $ 
induced from $ {\cal N}=2 $ super Schr\"odinger algebras $ {\cal G}_{(1,2,1)} $ and $ {\cal G}_{(2,2)}  $ which are defined on (1+1) dimensional spacetime. 
This was done based on the $D$-module presentations of $ {\cal G}_{(1,2,1)} $ and $ {\cal G}_{(2,2)}. $ However, the success of the generalization to \Z2-grading is due to the fact that $ {\cal G}_{(1,2,1)} $ and $ {\cal G}_{(2,2)} $ have an extra $ sl(2)$ doublet in addition to super symmetries $ (Q_a, S_a). $  
This is a common feature of super Schr\"odinger algebras. 
Therefore, more novel \Z2-graded superalgebras will be induced form higher $\cal N$ super Schr\"odinger algebras in 
arbitrary dimensional spacetime.

We also derived linear vector field realizations of $ \g^{(1,2,1)} $ and $ \g^{(2,2)} $ which will be useful for a generalization of superspace formulation to \Z2 setting. 
This is a first step to the full development of representation theory of \Z2-graded superalgebras induced from super Schr\"odinger algebras. 
What we really want to understand is whether \Z2-graded superalgebras are physically important algebraic structure on the same level as Lie algebras or Lie superalgebras. 
To this end we need to construct dynamical systems having \Z2-graded superalgebras as symmetry or dynamical algebras. 
There are various method to do it, for instance, Lie symmetry, nonlinear realization, coset construction and so on. 
Whichever method we take, understanding representations is a fundamental step to obtain dynamical systems. 
As a system which is related to \Z2-graded superalgebras we may possibly mention Dirac-Dunkle operators \cite{DD1,DD2,DD3}. 

  From mathematical point of view, representation theory of an algebraic structure is also very crucial. 
There exist many works on structure theory of the generalized (color) superalgebras \cite{sch2,sch3,SchZha,Sil,ChSiVO,CART} (and references therein). 
However, it seems that representation theory of them has not been investigated so well. 
The results of the present work pushed the representation theory forward and we wish to report further development in future publications.

%
%
%
\section*{Acknowledgements}

N. A. is supported by the  grants-in-aid from JSPS (Contract No. 26400209). 
J. S. would like to thank N. A. for invitation and hospitality at OPU. 

%
%
%
\setcounter{equation}{0}
\renewcommand{\theequation}{A.\arabic{equation}}
\section*{Appendix}

  In this appendix we construct a linear realization of the \Z2-graded superalgebra which is a generalization of 
${\cal N} = 1 $ super Schr\"odinger algebra. 

Restricting to $ a = 1 $ in the $D$-module representation of $ {\cal N} = 2 $ super Schr\"odinger algebra ${\cal G}_{1,2,1}$ (\ref{G121-D-mod}) 
one has its $ {\cal N} = 1 $ counterpart: 
\bea
  H &=& \i4 \del{t}, \nn \\
  D &=& -\i4 \Big(t\del{t} + \lambda + \hf x \del{x} \Big) -\hf(2e_{22} + e_{33} + e_{44}),
  \nn \\
  K &=& -\i4 t(t\del{t} + 2\lambda + x \del{x}) -t (2e_{22} + e_{33} + e_{44}),
  \nn \\
  Q &=& e_{13} + e_{42} + (e_{24}+e_{31}) \del{t},
  \nn \\
  S &=& t(e_{13} + e_{42}) + (e_{24}+e_{31}) (t \del{t} + 2\lambda + x \del{x}) + e_{24},
  \nn \\
  P &=& \i4 \del{x},
  \nn \\
  G &=& \i4 t \del{x},
  \nn \\
  X &=& (e_{24} + e_{31}) \del{x}. \label{N1-D-mod}
\eea
Their commutation relations are also $ a = 1 $ restriction of $ {\cal N} = 2 $ case.

Let us introduce new elements:
\beq
 \tP = \{P, P\},\quad U = \{P, G\}, \quad \tG = \{G, G\}, \quad \Pi=\{P, X\}, \quad \Lambda = \{G, X\}
\eeq
and \Z2-grading given by
\beq
   \begin{array}{ccl}
     (0,0) & : & H, \ D,\ K,\  \tP,\ U, \tG 
     \\[3pt]
     (0,1) & : & P,\ G
     \\[3pt]
     (1,0) & : & Q,\ S, \ \Pi, \ \Lambda
     \\[3pt]
     (1,1) & : & X
   \end{array}
\eeq
Then the (anti)commutators, which are computed with the aid of (\ref{N1-D-mod}), define a \Z2-graded superalgebra 
which we denote $ \gN1. $

 In order to introduce a triangular-like decomposition of $\gN1$, we take $D$ as a grading operator. 
The triangular-like dedomposition immediately follows from $D$-degree of the elements which reads  as follows:
\beq
   \begin{array}{ccl}
     +1 & : & H, \ \tP \\[3pt]
     +\hf & : & Q,\ \Pi,\ P \\[3pt]
     0 & : & D, \ U, \ X \\[3pt]
     -\hf & : & S, \ \Lambda, \ G \\[3pt]
     -1 & : & K, \ \tG
   \end{array}
\eeq

We parametrize a \Z2-graded group element for $ \gN1_+$ as
\beq
  g_+ = \exp(x_1 H) \exp(x_2 \tP) \exp(\psi P) \exp(\theta_1 Q) \exp(\theta_2 \Pi).
\eeq
Then the left action defined by (\ref{LAdef}) gives the linear realization of $\gN1.$

\medskip
\noindent
$\gN1_+ $ sector: 
\bea
   H &=& - \Del{x_1}, \nn \\
   \tP &=& - \Del{x_2}, \nn \\
   P &=& -\Del{\psi} + \hf \psi \Del{x_2},
   \nn \\
   Q &=& -\Del{\theta_1} + \theta_1 \Del{x_1},
   \nn \\
   \Pi &=& -\Del{\theta_2} + \theta_1 \Del{x_2},
\eea

\noindent
$\gN1_0$ sector:
\bea
   D &=& -\sum_{k=1}^2 \Big( x_k \Del{x_k} + \hf  \theta_k \Del{\theta_k} \Big) - \hf \psi \Del{\psi},
   \nn \\
   U &=& x_1 \Del{x_2} + \theta_1 \Del{\theta_2},
   \nn \\
   X &=& \theta_1 \Del{\psi} + \psi\Del{\theta_2} - \hf \psi \theta_1 \Del{x_2},
\eea

\noindent
$\gN1_-$ sector:
\bea
     \nn \\
   K &=& -2x_1 D - x_1^2 \Del{x_1} + 2x_2 \theta_1 \Del{\theta_2},
   \nn \\
   \tG &=& -x_1 U - x_1 \theta_1 \Del{\theta_2},
   \nn \\
   G &=& -x_1 P - \psi \theta_1 \Del{\theta_2},
   \nn \\
   S &=& -x_1 Q -  2x_2 \Pi - \theta_1 \Big( \psi \Del{\psi} + \theta_2 \Del{\theta_2}  \Big),
   \nn \\
   \Lambda &=& -x_1 \Pi.
\eea

%
%
%

\end{document}